# Spiral metasurface enables tunable directional edge enhancement

WENLI WANG, [1,] YAO HU, [2,3] QIUSHENG HUANG [1], DAN CHEN [1],

AND QUN HAO [*2,3]

*1 Chery Automobile Co., Ltd.*
*2 School of Optics and Photonics, Beijing Institute of Technology, Beijing 100081, China*
*3 National Key Laboratory on Near-Surface Detection, Beijing, 100072, China*
** Corresponding author: qhao@bit.edu.cn*

**Abstract:** Tunable directional edge enhancement facilitates the acquisition of distinct morphological features from objects, a capability that plays a vital role in enhancing the reliability and safety of autonomous driving systems. However, building a simple, miniature, and switchable directional edge enhancement system is still an urgent goal. To build this system, we present a compact spiral metasurface that not only is capable of producing tunable vertical and horizontal edge enhancement imaging, but also eliminates the need for a conventional 4f system by using only a single compact device. This is achieved by engineering the metasurface's spiral phase profile and its polarization-dependent response, where the direction of edge enhancement is controlled through variation of the incident beam's polarization, thus enabling switching between vertical and horizontal enhancement modes. Simulations confirm the metasurface's capability for switchable edge detection of lane markings and barrier contours. The broadband operation capability of the proposed metasurface is also verified by simulation. Owing to its compactness, switchable operation, and broadband performance, this spiral metasurface holds great promise for applications in optical analog computing and autonomous driving.

## 1. Introduction

Rapid and reliable directional edge enhancement is essential for autonomous driving, as it supports critical visual perception functions[1–3]. In particular, vertical edge enhancement sharpens the vertical alignment of lane markings, thereby enabling accurate vehicle positioning and lane keeping during straight driving, while horizontal enhancement focuses on identifying the contours of toll gate barriers to support obstacle detection at toll booths[4,5]. These one-dimensional (1D) two-directional enhancements work in tandem: their adaptive switching not only ensures centered driving by continuously locking onto lane markings on straight paths, but also quickly identifies lateral obstacles such as toll barriers and detects hazards from low vehicles or road signs during turns[5]. This dynamic dual-mode enhancement delivers irreplaceable technical value for autonomous driving. Hence, the combination and switching of these two edge enhancement capabilities significantly extend overall system reliability and safety.

To achieve such switchable edge enhancements, several optical-computation-based directional edge enhancement methods have been developed, including classical approaches such as the photonic spin Hall effect[6,7] and phase contrast imaging[8,9]. The implementation of these two techniques typically requires liquid crystal spatial light modulators, which facilitate switching between imaging modes via precise voltage control. Although such systems offer multifunctional imaging capabilities, their reliance on complex electro-optical modulation and multiple discrete elements results in systems that are less convenient than those based on simple incident beam mode regulation. Meanwhile, their bulky configurations and limited resolution hinder their applications in modern miniaturized and integrated systems, particularly where compact and efficient optical solutions are essential[10,11]. Thus, building a simple, miniature, high-resolution, switchable directional edge enhancement system has become a compelling and urgent goal for advanced autonomous driving applications.

In recent years, metasurfaces have emerged as a novel class of two-dimensional (2D) planar optical elements capable of replacing multiple bulky conventional optical devices. They provide precise phase control over light at sub-wavelength thicknesses, while also enabling flexible manipulation of both amplitude and polarization states of electromagnetic waves[12–15]. Furthermore, their ability to support switching between multiple optical functions makes them highly versatile for system integration[16–19]. These potentials have pushed metasurfaces into functional imaging areas[20–22], like 1D directional edge enhancement imaging[23,24], 2D anisotropic edge-enhancement imaging[25], and 2D isotropic edge enhancement imaging[26–28], which are all attractive. To the best of our knowledge, most existing metasurface-based edge enhancement imaging systems, especially those designed for 1D edge enhancement, continue to rely on a conventional 4f Fourier filtering architecture[20,29,30]. While this approach offers well-established advantages in spatial frequency manipulation, the requisite use of multiple macroscopic lenses introduces inherent optical complexity and considerable bulk[11,31]. Consequently, such implementations partially undermine the innate miniaturization potential and integration benefits of metasurfaces, suggesting a clear need for more compact and highly integrated system designs.

In this paper, we propose a compact spiral metasurface that can generate tunable vertical edge enhancement imaging and horizontal enhancement imaging by transforming the polarization of the incident beam. This metasurface is composed of many nanopillars, which are all prioritized to be appropriate for two independent phase profiles. Each profile combines a direction-controlled spiral phase with a parabolic phase. Consequently, the entire imaging process can be realized using only a single-layer metasurface, eliminating the need for a conventional 4f system. Additionally, the broadband operation capability of the proposed metasurface is verified.

## 2. Principle and Design

### *2.1 overall design*

Figure 1 is the schematic of our compact spiral metasurface system setup, consisting of only a helicity-multiplexing dielectric metasurface capable of dynamically switching between two different directional edge enhancement imaging by changing the polarization of the incident beam. For example, as shown in Fig.1(a), when the input object is a letter “E”, and the polarization of the incident beam is left-handed circular polarized (LCP), the output image is a vertical edge enhancement image of the object. Similarly, when this input object is under the illumination of a right-handed circular polarized (RCP) incident beam, the output information presented is a horizontal enhancement image of the object, as shown in Fig.1(b). In short, we can observe edge enhancement's directional edge enhancement characteristic (vertical or horizontal enhancement) from these. And other directional edge enhancements can also be tailored, which will be shown later.

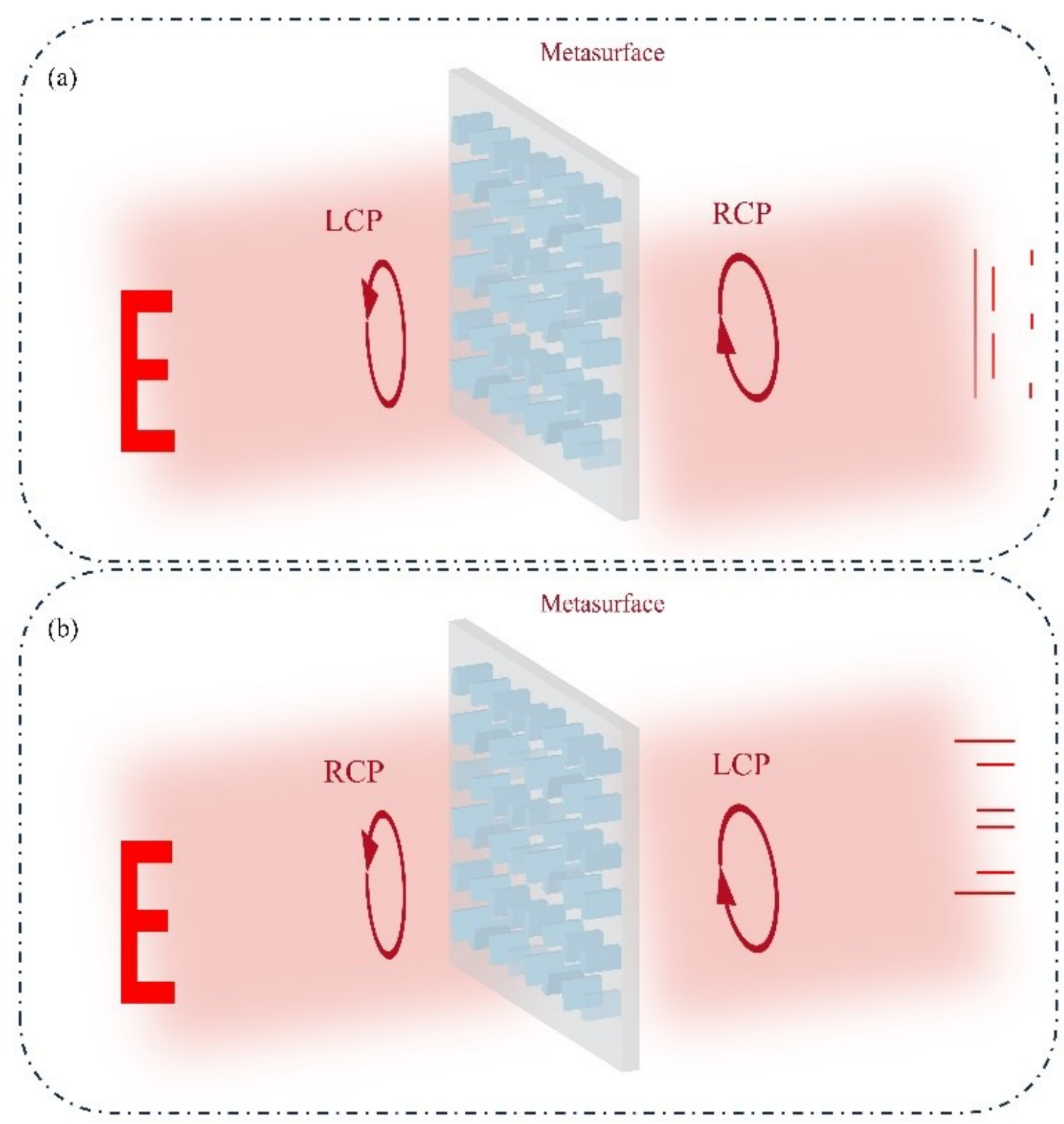

Fig.1 schematic of our compact spiral metasurface system setup. (a) a vertical edge enhancement image of the object "E" is obtained when illuminated by an LCP incident beam. (b) a horizontal enhancement image of the object "E" is obtained when illuminated by an RCP incident beam.

### *2.2 Principle of our compact spiral metasurface system*

To illustrate the feasibility of our compact spiral metasurface system, we analyze the whole optical image processing of this system and the relevant principle in detail. In this system, the object can be seen as an electric field pattern of the LCP or RCP incident beam written as $E_{in}$ $(x, y)$. $x$ and $y$ are respectively $x$ or $y$-direction coordinates in the input or output image plane. The point spread function of our system can be written as $PSF_s$. Therefore, the whole optical image processing can be concluded as the convolution of $E_{in}$ $(x, y)$ and $PSF_s$, following:

$$E_{out}\left(x,y\right) = E_{in}\left(x,y\right)\otimes \mathrm{PSF_s} \tag{1}$$

where the $PSF_s$ mainly rely on the phase distribution $\phi$ $(x, y)$ of our designed spiral metasurface, and their relationship can be written as :

$$\phi\left(x,y\right)\propto \mathrm{FT}[\mathrm{PSF_s}] \tag{2}$$

Next, one of the design keys is finding a proper $\phi$ $(x, y)$ to help us efficiently achieve directional edge enhancement imaging. Based on the demonstrated capability of directional edge enhancement imaging using a conventional 4f Fourier filtering system with a metasurface[32], we infer that the phase profile $\phi$ $(x, y)$ required for a single metasurface to achieve similar functionality should incorporate both the phase of a lens and the phase of the metasurface embedded within the 4f system. Thus, the phase of our compact spiral metasurface can be written as:

$$\phi(x,y) = -k_0[\sqrt{x^2+y^2+f_0^{\ 2}}-f_0)]+\mathrm{angle}(\exp(\mathrm{i}\varphi)+\exp(\mathrm{i}\varepsilon)\cdot\exp(-\mathrm{i}\varphi)), \tag{3}$$

where $f_0$ represents the designed focal length, angle (•) represents the operation of extracting the phase of the expression in parentheses. $\varphi$=arctan($y/x$). $\varepsilon$ is a real number. Several representative edge enhancements can be efficiently achieved by changing the value of $\varepsilon$.

To observe this characteristic clearly, we choose different values of $\varepsilon$, and then gain several different distributions of $\phi$ ($x$, $y$) by the angular spectrum method. Here, a circular pattern is used as the object $E_{in}$ ($x$, $y$). We can calculate corresponding output field patterns $E_{out}$ ($x$, $y$), as shown in Fig.2. Fig.2(a1) ~ (a5) displays the output field distributions when $\varepsilon$= 0, π/2, π, 3π/2, and 2π. The corresponding cross-sectional intensity curves are shown in Fig.2(b1) ~ (b5). It can be found that there were various edge enhancements along different directions when we changed the value of $\varepsilon$.

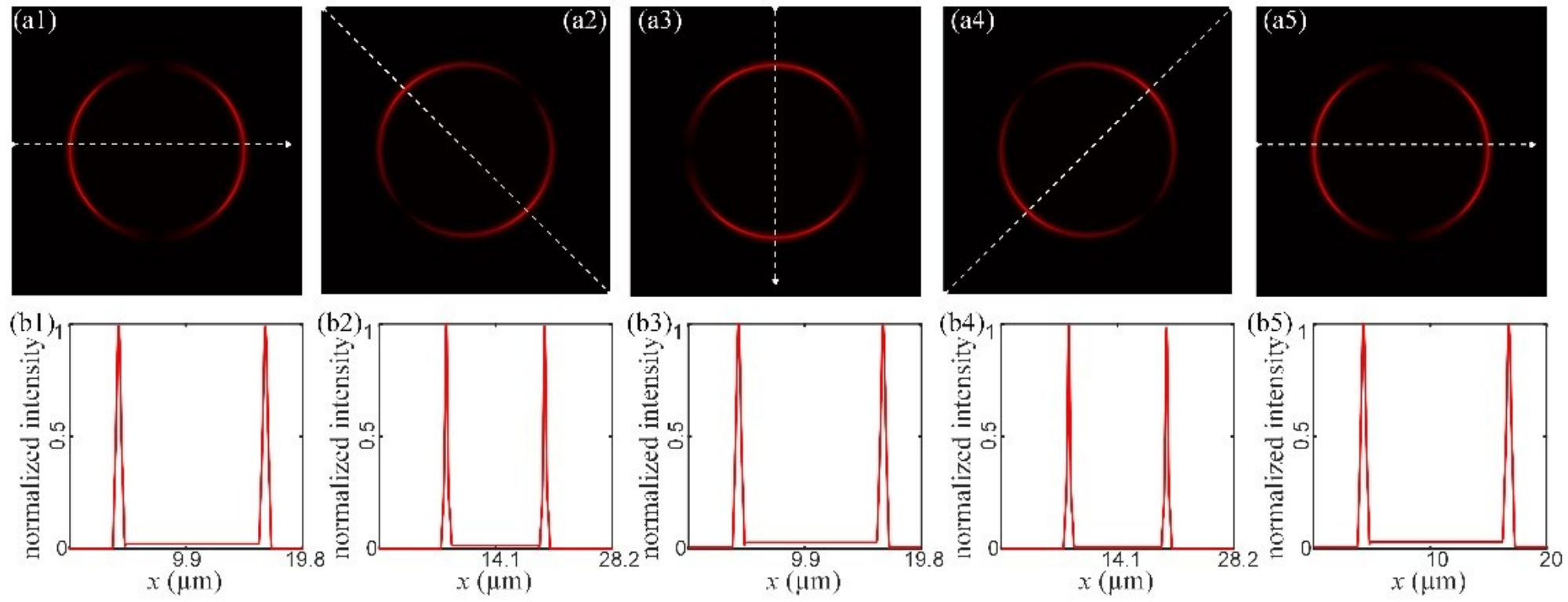

Fig.2 demonstrates the theoretical output results corresponding to the designed $\phi$ ($x$, $y$) when a circular pattern is the input object. (a1) ~ (a5) are theoretical output results as we set the $\varepsilon$= 0, π/2, π, 3π/2 and 2π, respectively. (b1) ~ (b5) are the cross-sectional intensity curve of (a1) ~ (a5) along the white dashed line.

Therefore, the required phase profiles for our compact spiral metasurface system mentioned in Fig.1 can be written like this:

$$\phi_1(x,y) = -k_0[\sqrt{x^2+y^2+f_0^2} - f_0)] + \text{angle}(\exp(\mathrm{i}\varphi) + \exp(\mathrm{i}\pi)\cdot\exp(-\mathrm{i}\varphi)), \tag{4}$$

$$\phi_2(x,y) = -k_0[\sqrt{x^2+y^2+f_0^2} - f_0)] + \text{angle}(\exp(\mathrm{i}\varphi) + \exp(\mathrm{i}0)\cdot\exp(-\mathrm{i}\varphi)), \tag{5}$$

where, $\phi_1$ and $\phi_2$ are respectively the required phase profiles when the polarization of the incident beam is LCP or RCP.

Then, we divide the total phase of our metasurface into the propagation phase and the geometry phase, which can be respectively expressed with the phase profiles in Eq. (4) ~ (5) [30].

$$\delta_x(x,y) = \left[\phi_1(x,y) + \phi_2(x,y)\right]/2, \tag{6}$$

$$\delta_y(x,y) = \left[\phi_1(x,y) + \phi_2(x,y)\right]/2 - \pi, \tag{7}$$

$$2\theta(x,y) = \left[\phi_1(x,y) - \phi_2(x,y)\right]/2. \tag{8}$$

where, $\delta_x$ and $\delta_y$ are the propagation phase distributions for a linear polarization incident beam in $x$-and $y$-directions, respectively; $2\theta$ is the geometric phase distribution.

### *2.3 Design of our compact spiral metasurface*

As illustrated in Fig. 3(a), which shows the side view, a specifically designed unit cell made of titanium dioxide nanopillars on a glass substrate conforms to the required propagation and geometric phase. These nanopillars are arranged in a periodic square lattice with fixed constants $P_x$=$P_y$=360nm nm and a height $H$ = 600 nm[25]. The propagation phase library is constructed by varying the nanopillar's length $L$ and width $W$. To cover 0~2π, the range of $L$ and $W$ of the nanobricks covers 76 to 282nm, with 2 nm increments of each geometric variable. The schematic for computing the transmission coefficients ($t_x$, $t_y$) and phase shifts ($\delta_x$, $\delta_y$) is shown in Fig. 3b. Figure 3(c) displays the simulated phase shift $\delta_x$ for $x$-direction linearly polarized (XLP) incident light as a function of $L$ and $W$, with the corresponding transmission amplitude $t_x$ shown in Fig. 3(d). Similarly, Fig. 3(e) and (f) present the simulated phase shift $\delta_\gamma$ and transmission amplitude $t_\gamma$ for $y$-direction linearly polarized (YLP) incidence. The geometric phase is controlled by rotating the nanopillar's orientation angle $\theta$, producing a phase shift equal to $2\theta$ that readily covers the full phase range. All nanopillars are designed to function as half-wave plates for maximal polarization conversion efficiency. Through appropriate selection of the parameters $L$, $W$, and $\theta$, any desired phase combination ($\delta_x$, $\delta_\gamma$, $2\theta$) can be achieved, thereby enabling the design of the metasurface that meets the phase profiles given in Eq. (4) ~ (5). The simulations were conducted using the finite-difference time-domain (FDTD) method under plane-wave illumination at a wavelength of 532 nm.

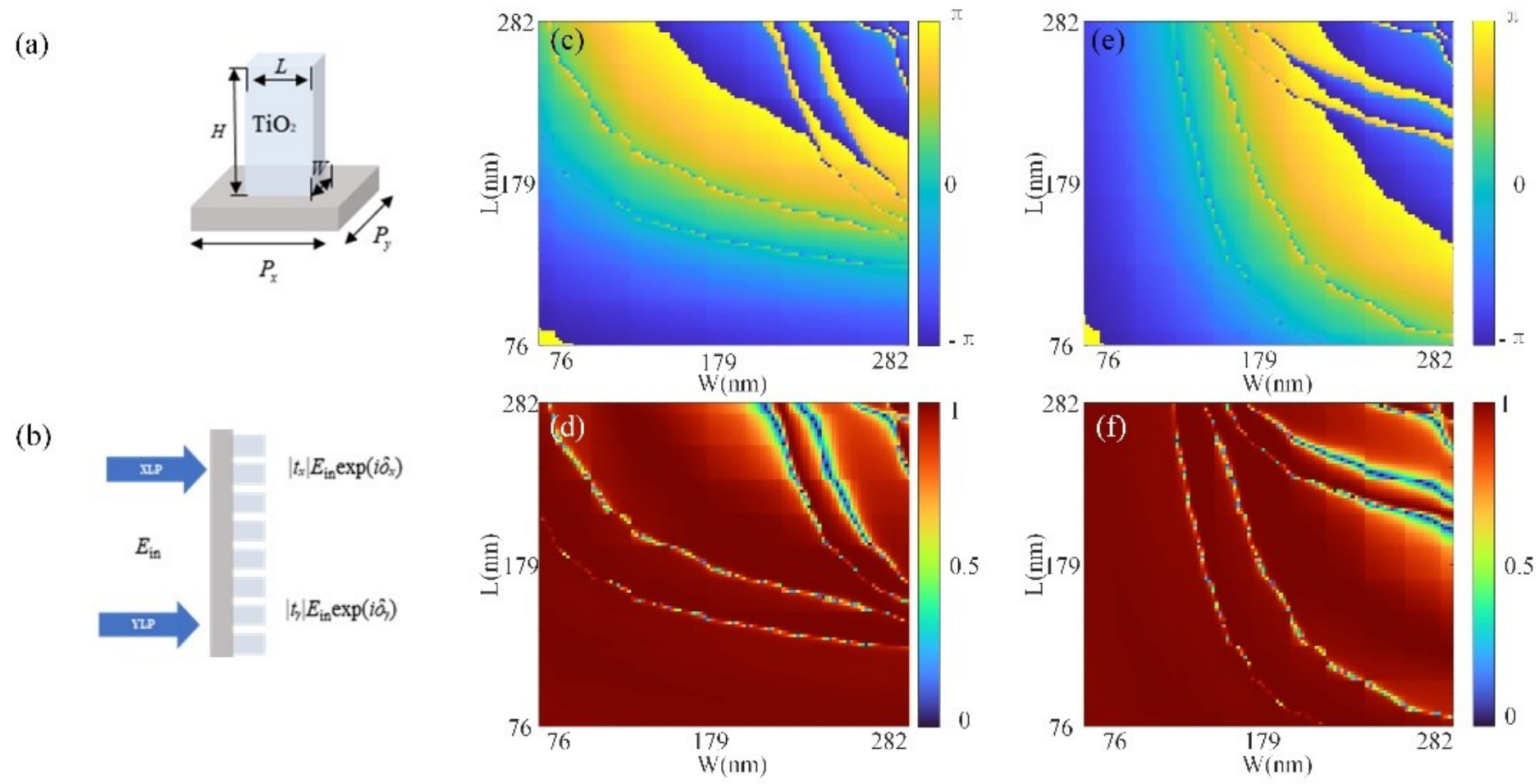

Fig.3 (a) is a side view of a typical unit cell with the period ($P_x$, $P_y$), height (H), varying cross sizes ($L$ and $W$), and varying orientation angle ($\theta$). (b) is schematic for the numerical calculation of the transmission coefficients ($t_x$, $t_y$) and phase shifts ($\delta_x$, $\delta_\gamma$). (c,d) are simulated propagation phase $\delta_x$ and transmission coefficient distribution $t_x$, for an XLP incident beam. (e,f) are simulated propagation phase $\delta_\gamma$ and transmission coefficient distribution $t_y$ for a YLP incident beam.

## 3. Results

To comprehensively demonstrate the feasibility of our spiral metasurface system for optical edge detection, we simulated the entire imaging process using the angular spectrum propagation method. The phase profile of the metasurface was designed and verified using the FDTD method. The systematic edge-enhancement capability of the system is presented in Fig. 4. It shows the results for vertical and horizontal edges under different incident circular polarizations, using three distinct test objects: the letter "E", an image of a lane marking, and a picture of a boom barrier. These examples collectively demonstrate the generality of our approach.

Consistent with the design, under the illumination of LCP, our system selectively enhances vertical edges, as clearly displayed by the processed output images in Fig. 4(a1) ~ Fig. 4(a3). The intensity profiles extracted along the white dashed lines in these pictures are presented in Fig. 4(b1) ~ Fig. 4(b3), respectively. The pronounced peaks in these profiles, which correspond precisely to the vertical edges in the original images, confirm the targeted edge enhancement. Conversely, when the incident beam is RCP, the system switches its function to highlight horizontal edges, as shown in the resulting images of Fig. 4(c1) ~ Fig. 4(c3). The intensity profiles extracted along the white dashed lines in these figures are respectively presented in Fig.

4(d1) ~ Fig. 4(d3), respectively. The pronounced peaks in these profiles, which correspond to the horizontal edges in the original images, confirm the targeted edge enhancement. The consistent and robust performance across all three distinct test objects underscores the generality of the image-enhancement functionality.

In conclusion, these results shown in Fig. 4 successfully verify that our spiral metasurface system achieves switchable, directional edge detection in images, with the operating mode (vertical or horizontal edge enhancement) controlled simply by altering the incident polarization state.

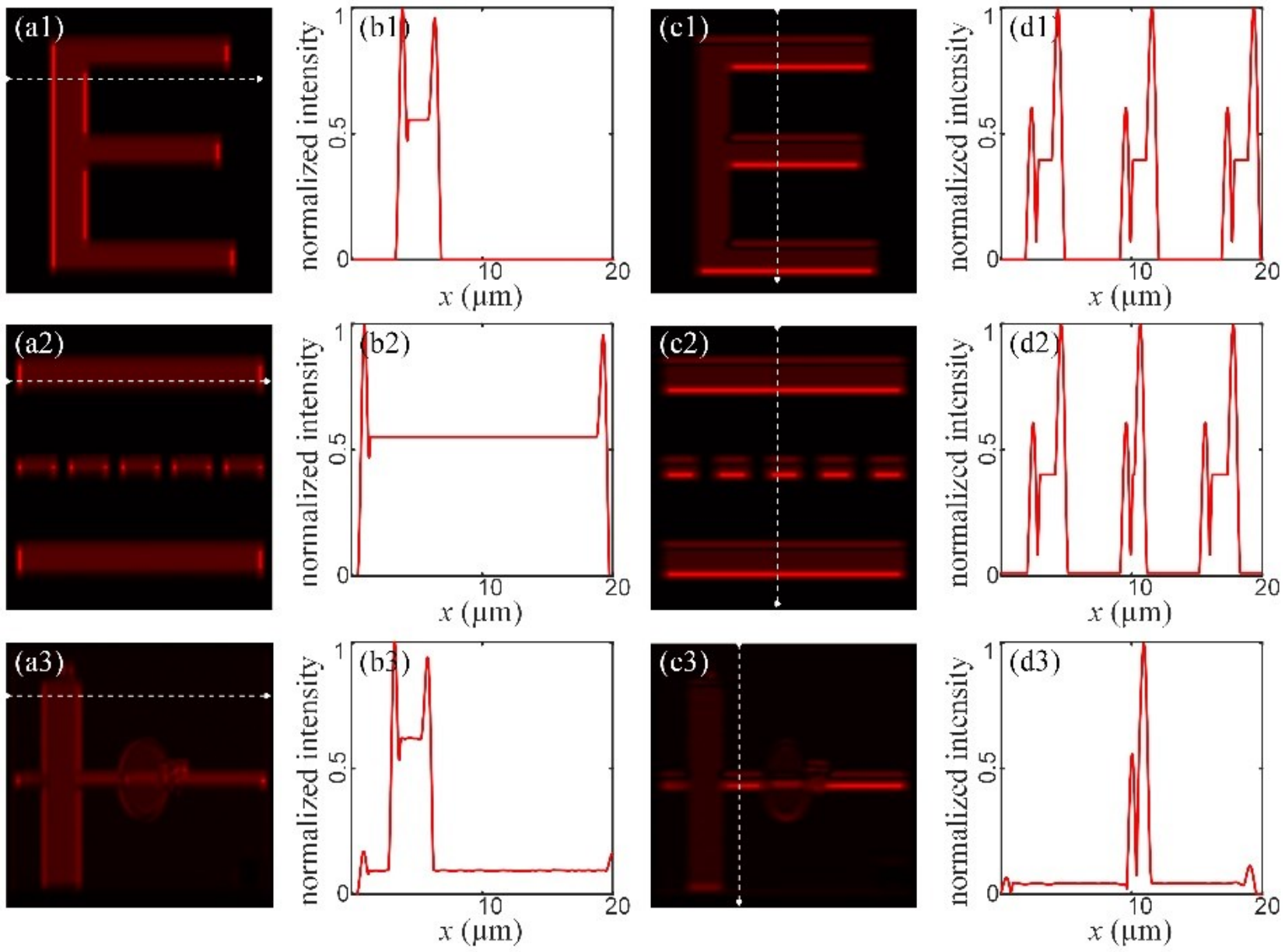

Fig.4 (a1) ~ (a3) respectively show the processed output images when the incident beam is LCP and the test objects are a letter "E", an image of a lane marking, and a picture of a boom barrier. (b1) ~ (b3) are intensity profiles shown by the white dashed lines in (a1) ~ (a3). (c1) ~ (c3) respectively show the processed output images when the incident beam is RCP and the test objects are respectively a letter "E", an image of a lane marking, and a picture of a boom barrier. (d1) ~ (d3) are intensity profiles shown by the white dashed lines in (c1) ~ (c3).

Since the geometric phase can operate over a wide bandwidth, we infer that our spiral metasurface system, which is based on a combination of the geometric phase and the propagation phase, should also function across a broad band. To verify the broadband performance of our system, we selected the letter "E" as a test object. The letter "E" was illuminated with distinct incident beams of wavelengths 520 nm, 532 nm, and 540 nm. The corresponding simulation results are presented in Fig. 5.

Figure 5(a1) ~ (a3) shows the vertical enhancement capability of our system for incident beam wavelengths of 520 nm, 532 nm, and 540 nm, respectively, under LCP illumination. The corresponding intensity profiles, extracted along the white dashed lines in each image, are shown in Figs. 5(b1) ~ (b3). The pronounced peaks in these intensity profiles, which align with the vertical edges in the original images, confirm the broadband edge enhancement capability. Moreover, Figure 5(c1) ~ (c3) demonstrates the horizontal edge enhancement capability of our system under RCP illumination across wavelengths of 520, 532, and 540 nm. The intensity profiles in Figs. 5(d1) ~ (d3), extracted along the white dashed lines, exhibit pronounced peaks that align with the horizontal edges in the original images, confirming the broadband performance.

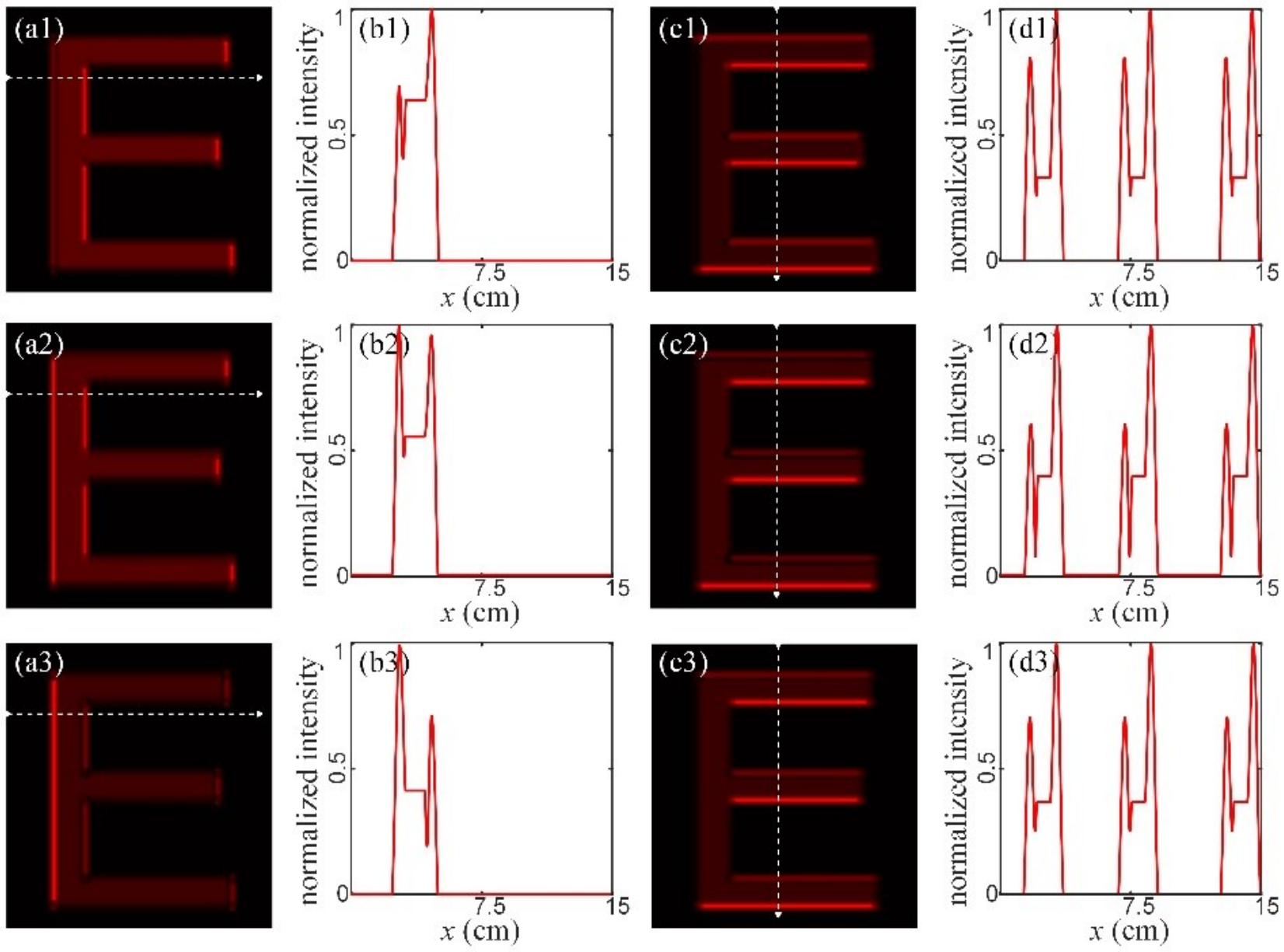

Fig.5 (a1) ~ (a3) shows the processed output images under LCP illumination for incident beam wavelengths of 520 nm, 532 nm, and 540 nm, respectively. (b1) ~ (b3) display the intensity profiles extracted along the white dashed lines in (a1) ~ (a3). (c1) ~ (c3) show the processed output images under RCP illumination for incident beam wavelengths of 520 nm, 532 nm, and 540 nm, respectively. (d1) ~ (d3) display the intensity profiles extracted along the white dashed lines in (a1) ~ (a3).

## 4. Conclusion

In conclusion, we have developed a compact spiral metasurface that realizes tunable vertical and horizontal edge enhancement imaging by simply controlling the polarization of the incident beam, thereby achieving the desired imaging performance without relying on a conventional 4f optical system. To achieve these functions, our metasurface is designed to support two independent phase profiles, each integrating a direction-controlled spiral phase with a parabolic phase. We have demonstrated that our metasurface can achieve switchable edge detection of lane markings and barrier contours. The broadband operation capability is also verified. Benefiting from the advantages of a planar architecture and the broadband, dynamically switchable functionality of the metasurface, we foresee that this spiral design offers significant potential for applications in optical analog computing and autonomous driving.

**Data availability**

The data that support the plots within this paper and other findings of this study are available from the authors upon reasonable request.

**Conflicts of interest**

There are no conflicts to declare.

**Acknowledgments**

This work was supported by the National Key R&D Program of China (2023YFF0718101) and the National Natural Science Foundation of China (62427817).

### Reference

1. L. Li, J. Ma, D. Sun, Z. Tian, L. Cao, and P. Su, "Amp-vortex edge-camera: a lensless multi-modality imaging system with edge enhancement," Opt. Express **31**(14), 2519-22531 (2023).

2. C. Chen, C. Wang, B. Liu, C. He, L. Cong, and S. Wan, "Edge Intelligence Empowered Vehicle Detection and Image Segmentation for Autonomous Vehicles," IEEE Trans. Intell. Transp. Syst. **24**(11), 13023–13034 (2023).
3. X. Li, T. Zhang, S. Wang, G. Zhu, R. Wang, and T.-H. Chang, "Large-Scale Bandwidth and Power Optimization for Multi-Modal Edge Intelligence Autonomous Driving," IEEE Wirel. Commun. Lett. **12**(6), 1096–1100 (2023).
4. S. Liu, L. Liu, J. Tang, B. Yu, Y. Wang, and W. Shi, "Edge Computing for Autonomous Driving: Opportunities and Challenges," Proc. IEEE **107**(8), 1697–1716 (2019).
5. W. Wang, Y. Hu, Q. Huang, and Q. Hao, "Polarization metasurface enables tunable 1D directional edge enhancement," in *Advanced Optical Imaging Technologies VIII*, P. S. Carney, X.-C. Yuan, and K. Shi, eds. (SPIE, 2025), p. 43.
6. Y. Wang, Q. Yang, S. He, R. Wang, and H. Luo, "Computing Metasurfaces Enabled Broad-Band Vectorial Differential Interference Contrast Microscopy," ACS Photonics.**10**(7), 2201–2207 (2022).
7. T. Zhu, Y. Lou, Y. Zhou, J. Zhang, J. Huang, Y. Li, H. Luo, S. Wen, S. Zhu, Q. Gong, M. Qiu, and Z. Ruan, "Generalized Spatial Differentiation from the Spin Hall Effect of Light and Its Application in Image Processing of Edge Detection," Phys. Rev. Appl. **11**(3), 034043 (2019).
8. F. Zernike, "How I Discovered Phase Contrast," Science. **121**(3141), 345–349 (1955).
9. M. Ritsch-Marte, "Orbital angular momentum light in microscopy," Philos. Trans. R. Soc. A Math. Phys. Eng. Sci. **375**(2087), 20150437 (2017).
10. C. Maurer, A. Jesacher, S. Bernet, and M. Ritsch-Marte, "What spatial light modulators can do for optical microscopy," Laser Photon. Rev. **5**(1), 81–101 (2011).
11. Y. Zhang, P. Lin, P. Huo, M. Liu, Y. Ren, S. Zhang, Q. Zhou, Y. Wang, Y. Lu, and T. Xu, "Dielectric Metasurface for Synchronously Spiral Phase Contrast and Bright-Field Imaging," Nano Lett. **23**(7), 2991–2997 (2023).
12. A. I. Kuznetsov, M. L. Brongersma, J. Yao, M. K. Chen, U. Levy, D. P. Tsai, N. I. Zheludev, A. Faraon, A. Arbabi, N. Yu, D. Chanda, K. B. Crozier, A. V. Kildishev, H. Wang, J. K. W. Yang, J. G. Valentine, P. Genevet, J. A. Fan, O. D. Miller, A. Majumdar, J. E. Fröch, D. Brady, F. Heide, A. Veeraraghavan, N. Engheta, A. Alù, A. Polman, H. A. Atwater, P. Thureja, R. Paniagua-Dominguez, S. T. Ha, A. I. Barreda, J. A. Schuller, I. Staude, G. Grinblat, Y. Kivshar, S. Peana, S. F. Yelin, A. Senichev, V. M. Shalaev, S. Saha, A. Boltasseva, J. Rho, D. K. Oh, J. Kim, J. Park, R. Devlin, and R. A. Pala, "Roadmap for Optical Metasurfaces," ACS Photonics **11**(3), 816–865 (2024).
13. Y. Yang, E. Lee, Y. Park, J. Seong, H. Kim, H. Kang, D. Kang, D. Han, and J. Rho, "The Road to Commercializing Optical Metasurfaces: Current Challenges and Future Directions," ACS Nano **19**(3), 3008–3018 (2025).
14. N. A. Rubin, P. Chevalier, M. Juhl, M. Tamagnone, R. Chipman, and F. Capasso, "Imaging polarimetry through metasurface polarization gratings," Opt. Express **30**(6), 9389-9412 (2022).
15. S. Li, C. Chen, G. Wang, S. Ge, J. Zhao, X. Ming, W. Zhao, T. Li, and W. Zhang, "Metasurface Polarization Optics: Phase Manipulation for Arbitrary Polarization Conversion Condition," Phys. Rev. Lett. **134**(2), 023803 (2025).
16. L. Zhang, Z. Zhao, L. Tao, Y. Wang, C. Zhang, J. Yang, Y. Jiang, H. Duan, X. Zhao, S. Chen, and Z. Wang, "A Review of Cascaded Metasurfaces for Advanced Integrated Devices," Micromachines **15**(12), 1482 (2024).
17. Y. Yang, J. Seong, M. Choi, J. Park, G. Kim, H. Kim, J. Jeong, C. Jung, J. Kim, G. Jeon, K. Lee, D. H. Yoon, and J. Rho, "Integrated metasurfaces for re-envisioning a near-future disruptive optical platform," Light Sci. Appl. **12**(1), 152 (2023).
18. Z. Wang, Y. Xiao, K. Liao, T. Li, H. Song, H. Chen, S. M. Z. Uddin, D. Mao, F. Wang, Z. Zhou, B. Yuan, W. Jiang, N. K. Fontaine, A. Agrawal, A. E. Willner, X. Hu, and T.

Gu, "Metasurface on integrated photonic platform: from mode converters to machine learning," Nanophotonics **11**(16), 3531–3546 (2022).
19. X. Wang, C. Wang, J. Ni, S. Liu, H. Wu, Y. Tao, Y. Jin, X. Wang, J. Li, Y. Hu, J. Chu, S. Chen, H. Ren, and D. Wu, "Ultracompact Achromatic Spiral Phase Contrast Imager on a CMOS Chip," Laser Photon. Rev. **19**(18), e00239 (2025).
20. M. Cotrufo, S. Singh, A. Arora, A. Majewski, and A. Alù, "Polarization imaging and edge detection with image-processing metasurfaces," Optica **10**(10), 1331-1338 (2023).
21. R. Wang, S. He, S. Chen, W. Shu, S. Wen, and H. Luo, "Computing metasurfaces enabled chiral edge image sensing," iScience **25**(7), 104532 (2022).
22. Y. Lian, Y. Liu, D. Cheng, C. Chi, Y. Bao, and Y. Wang, "Dual-mode varifocal Moiré metalens for quantitative phase and edge-enhanced imaging," Nanophotonics **14**(18), 3053–3062 (2025).
23. J. Sardana, S. Devinder, S. Kaassamani, W. Zhu, A. Agrawal, and J. Joseph, "Polarization Specific Edge Enhancement Enabled by Compact Dielectric Metasurface Imaging System," ACS Photonics **12**(5), 2380–2388 (2025).
24. J. Zhou, S. Liu, H. Qian, Y. Li, H. Luo, S. Wen, Z. Zhou, G. Guo, B. Shi, and Z. Liu, "Metasurface enabled quantum edge detection," Sci. Adv. **6**(51), eabc4385 (2020).
25. Q. Hao, W. Wang, J. Wang, Q. Li, Y. Hu, S. Zhang, and L. Yu, "Computing metasurface capable of broad-band switchable anisotropic edge-enhanced imaging," J. Mater. Chem. C **11**(12), 3956–3963 (2023).
26. T. Badloe, Y. Kim, J. Kim, H. Park, A. Barulin, Y. N. Diep, H. Cho, W.-S. Kim, Y.-K. Kim, I. Kim, and J. Rho, "Bright-Field and Edge-Enhanced Imaging Using an Electrically Tunable Dual-Mode Metalens," ACS Nano **17**(15), 14678–14685 (2023).
27. P. Zhou, K. Guo, Q. Ji, X. Wen, C. Guo, H. Zheng, R. Guo, L. Chen, H. Ye, Y. Liu, and S. Deng, "Full-Stokes Polarization Bifocal Plane Metasurface with Bright-Field and Edge-Enhanced Imaging," Laser Photon. Rev, e00922 (2025)[ Early Access].
28. H. Yang, Z. Xie, H. He, Q. Zhang, J. Li, Y. Zhang, and X. Yuan, "Switchable imaging between edge-enhanced and bright-field based on a phase-change metasurface," Opt. Lett. **46**(15), 3741-3744 (2021).
29. J. Zhou, H. Qian, C.-F. Chen, J. Zhao, G. Li, Q. Wu, H. Luo, S. Wen, and Z. Liu, "Optical edge detection based on high-efficiency dielectric metasurface," Proc. Natl. Acad. Sci. **116**(23), 11137–11140 (2019).
30. P. Huo, C. Zhang, W. Zhu, M. Liu, S. Zhang, S. Zhang, L. Chen, H. J. Lezec, A. Agrawal, Y. Lu, and T. Xu, "Photonic Spin-Multiplexing Metasurface for Switchable Spiral Phase Contrast Imaging," Nano Lett. **20**(4), 2791–2798 (2020).
31. Y. Kim, G. Lee, J. Sung, J. Jang, and B. Lee, "Spiral Metalens for Phase Contrast Imaging," Adv. Funct. Mater. **32**(5), 2106050 (2022).
32. M. K. Sharma, J. Joseph, and P. Senthilkumaran, "Directional edge enhancement using superposed vortex filter," Opt. Laser Technol. **57**, 230–235 (2014).